\RequirePackage{fix-cm}
\documentclass[smallextended]{svjour3}       % onecolumn (second format)
\smartqed  % flush right qed marks, e.g. at end of proof
\usepackage{graphicx}
\usepackage{float}
\usepackage{hyperref}
\usepackage{xcolor}
\usepackage{comment}
\usepackage{slashed} 
\usepackage{amsmath}
\usepackage{autobreak}
\usepackage{hyperref}

\begin{document}

\title{Mass spectroscopy of charmonium using a screened potential}


\author{Sreelakshmi M         \and
        Akhilesh Ranjan %etc.
}


\institute{Sreelakshmi M \at
             Department of Physics, Manipal Institute of Technology\\ Manipal 
Academy of Higher Education, Manipal, 576104, Karnataka, India\\
              \email{sreelakshmim.araam@gmail.com}           %  \\
           \and
           A. Ranjan \at
               Department of Physics, Manipal Institute of Technology\\ Manipal 
Academy of Higher Education, Manipal, 576104, Karnataka, India\\
 \email{ak.ranjan@manipal.edu}    
}

\date{Received: date / Accepted: date}

\maketitle

\begin{abstract}
In this work, we estimate the mass spectra and decay properties of charmonium 
($c \bar c$) using a non-relativistic potential model. We employ a potential 
model incorporating a Coulomb like term, representing one gluon exchange at 
short distances, and a screening term representing quark confinement at long 
distances. Spin-dependent corrections are also added perturbatively. Our results are compared with available experimental data and 
some other theoretical models. Based on this, we have made some comments on 
interquark potential.

\end{abstract}
%\newpage
\section{Introduction} \label{introduction}
 
\label{intro}

There are various experimental facilities across the world, like LHCb \cite{lhc}, Belle \cite{choi}, CLEO \cite{cleo}, and BaBar \cite{babar} etc that investigate the properties of hadrons. Hadrons are composed of quarks and antiquarks and follow the principles of Quantum Chromodynamics (QCD). QCD is the theory of the strong interaction. Studying the properties of hadrons provides deep insight into the strong interaction. Gluons are the mediators of the strong interaction between quarks and also carry color charges. 

One unique feature of strong interaction is asymptotic freedom. According to 
this, the effective coupling constant ($\alpha_{s}=g_{s}/4\pi$) decreases 
logarithmically at short distances, which is also well supported theoretically \cite{gross}. Another special feature of strong interaction is color confinement. 
According to color confinement, only color singlet hadrons can freely exist in 
nature. Color confinement property \cite{wilson}, is supported by lattice 
simulations but not yet theoretically proven. Any strong interaction theory 
should incorporate these two features. There are many theoretical methods 
used in hadron spectroscopy. Some of the highly used methods are the bag model 
\cite{chodos}\cite{chodos1}, QCD sum rules \cite{shifman}, Heavy Quark 
Effective Theory (HQET) \cite{korner}, phenomenological potential models 
\cite{eichten}\cite{martin}\cite{bhanot}, Lattice QCD (LQCD) including lattice 
gauge theories \cite{kogut}, etc.

Phenomenological potential models have been highly successful in describing the 
properties of hadrons, including both light and heavy quark systems. These 
models are inspired by the characteristics of QCD, confinement, and asymptotic 
freedom. For interactions at shorter distances, one gluon exchange (OGE) dominates, 
while at large interquark distances, QCD perturbation theory breaks down, and quarks 
become confined. The Cornell potential is a widely used phenomenological potential 
that combines a Coulomb-like term to account for one gluon exchange and a linear 
term to describe quark confinement \cite{eichten}\cite{kuchin}. These models often 
provide remarkably accurate descriptions of experimental data, making them a 
significant method for investigating the properties of hadrons. In the case of 
heavy hadrons, the potential can be treated non-relativistically and solved using 
the Schr\"{o}dinger equation. Therefore, in this work, we have used a 
phenomenological potential model.

The study of quarkonium ($q \bar q$) has been very helpful in understanding 
the strong interaction after the discovery of $J/\psi$ \cite{augustin} at SLAC. 
Quarkonium is relatively simple to study, as it consists of one quark 
and its antiquark, which is a two-body system. We are finding many interesting experimental results on exotic multiquark states these days \cite{abazov}\cite{aaltonen}. The studies on quarkonium can be 
the base for understanding such exotic multiquark states
like tetraquarks and pentaquarks. This also makes the study of quarkonium systems 
important.

In this paper, we have estimated the mass and decay properties of charmonium using 
a non-relativistic screened potential. The paper is organized as follows: a brief 
introduction in Section \ref{introduction} is followed by the theoretical 
formulation of the problem in Section \ref{formulation}. Discussion of the results 
in Section \ref{discussion} is followed by conclusions in Section \ref{conclusion}.

\section{Theoretical Formulation} \label{formulation}
The QCD is a non-abelian and complex theory. Therefore, the real nature of the 
interquark interaction is still unknown. The QCD vacuum contains virtual 
quark-antiquark pairs and gluons. A quark inside a hadron attracts an opposite 
color charge, and a cloud of virtual particles builds up around the quark. This, 
in turn, leads to a screening effect on other quarks at a distance. Therefore,
it is worth using screened potential to study quark interactions.
We have used a non-relativistic screened potential for the analysis of quarkonia. 
A screened potential $(V_{NR}(r))$ is a combination of the Coulomb term and screening term \cite{lucha1}. Apart from the non-relativistic contribution ($V_{NR}(r)$), interaction potential includes spin-dependent correction ($V_{SD}(r)$), which is added perturbatively to the Hamiltonian,

\begin{equation}
   V(r)=V_{NR}(r)+V_{SD}(r).
 \end{equation}
 
 The form of screened potential can be written as \cite{patel1},

\begin{equation}
V_{NR}(r)= \frac{k_{s}\alpha_{s}}{r}+b\left(\frac{1-\exp^{(-\mu_{s} r)}}{\mu_{s}}\right)-c,
\end{equation}

where $b$ and $c$ are potential parameters, $\alpha_{s}$ is the strong coupling constant and $\mu_{s}$ is the parameter for screening effect. The parameter $k_{s}$ represents the color factor and denotes a color singlet state. From $SU(3)$ representation, $3\otimes \bar 3= 1\oplus8$, leads to $-\frac{4}{3}$ for a color singlet state.

A screened potential introduces light quark-antiquark pairs in the vacuum, which softens the interaction at large distances compared to the linear potential. At short distances, $r \leq 1 fm$, the screened potential behaves identically to a linear potential.  Screening lowers the mass of higher excited states \cite{patel1}. 

In the case of heavy hadrons, the kinetic energy of the quarks is less compared to their rest mass energy. Therefore, a non-relativistic approach can be used. The time-independent radial Schr\"{o}dinger equation is, 

\begin{equation}
%\[\left(\frac{1}{2}\right) \qquad (\frac{1}{2})\]
\left[\frac{1}{2\mu}\left(-\frac{d^{2}}{dr^{2}}\right)+\frac{l(l+1)}{r^{2}}+V(r)\right]y(r)=Ey(r),
\end{equation}

parameter $\mu$ is the reduced mass, $\mu=\frac{m_{1}m_{2}}{m_{1}+m_{2}}$. The Schr\"{o}dinger equation is solved by numerical integration with the Runge-Kutta method using Mathematica package \cite{lucha2}. We have fitted the center of weight mass or spin average mass ($M_{SA_{,nJ}}$) of the ground state with the known experimental values (from Particle Data Group \cite{pdg}) of pseudoscalar and vector mesons, 
            \begin{equation}
             M_{SA_{,nJ}}=\frac{\sum_{J}(2J+1)M_{nJ}}{\sum_{J}(2J+1)}.
         \end{equation}
         The constant parameters used for the calculations are given in Table \ref{table:1}. Spin-dependent terms will be considered perturbatively. The correction contains spin-spin, spin-orbit, and tensor interactions. The spin-spin interaction creates hyperfine levels, whereas spin-orbit and tensor interactions are responsible for the fine structure of the quarkonium. The coefficients of these spin-dependent terms can be expressed as \cite{lucha1},

%\begin{autobreak}
{\small
\begin{equation}
 V_{SD}(r)=V_{SS}(r)\left[S(S+1)-\frac{3}{2}\right]+V_{L \cdot S}(r)(\mathbf{L \cdot S})+V_{T}(r)\left[S(S+1)-3\frac{(\mathbf{S \cdot r )(S \cdot r)}}{r^{2}}\right],
\end{equation}}
%\end{autobreak}

where $S$ is the total spin and $L$ is total orbital angular momentum. These terms 
are the result of Lorentz structure ($\Gamma_{\Sigma} \otimes \Gamma_{\Sigma }$) of 
inter-quark interaction. Various Lorentz structures of a fermion-anti fermion interaction will give rise to scalar ($V_{s}(r)$), pseudo scalar (zero), vector  ($V_{v}(r)$), axial vector ($V_{A}(r)$) and tensor ($V_{T}(r)$) terms in a non-relativistic potential \cite{lucha3}. Therefore, the effective potential should be considered as the sum of vector and scalar terms,
\begin{equation}
    V(r)=V_{v}(r)+V_{s}(r).
\end{equation}

%The spin-dependent part added to the one gluon exchange potential is considered to obtain the mass difference between degenerate mesonic states.

The spin-spin term for equal quark masses is given by,

\begin{equation}V_{SS}(r)=\frac{32 \pi \alpha_{s}}{9 m_{Q}^{2}}|\psi(0)|^{2}\langle \mathbf{S_{1} \cdot S_{2}} \rangle,\end{equation}

where $m_{Q}$ is heavy quark mass and $S_{1}$ and $S_{2}$ are the spin of the heavy quark and heavy antiquark, respectively, with,

\begin{equation}
\langle \mathbf{S_{1} \cdot S_{2}} \rangle=
    \begin{cases}
         -\frac{3}{4}  &   S=0, \textrm {spin singlets}\\
         \frac{1}{4}& S=1, \textrm{spin triplets}.
    \end{cases}
\end{equation}

The spin-orbit term is given by,

\begin{equation}V_{L \cdot S}(r)=\frac{\mathbf{L \cdot S}}{(2 m_{Q})^{2}r} \left[\frac{1}{r}\frac{dV_{v}(r)}{dr}-\frac{d^{2}V_{v}(r)}{dr^{2}}\right],\end{equation}

where $L$ and $S$ are the total orbital angular momentum and total spin of the heavy 
quark and heavy antiquark, respectively, with,

\begin{equation}
\left\langle L \cdot S \right\rangle=
    \begin{cases}
        l, &   J=l+1\\
       -1, & J=l\\
       -(l+1),  &  J=l-1.
    \end{cases}
\end{equation}

The tensor term is given by,

\begin{equation}V_{T}(r)=\frac{\mathbf{S_{12}}}{(12m_{Q})^{2}} \left[3\frac{dV_{v}(r)}{dr}-\frac{dV_{s}(r)}{dr}\right],\end{equation}

\begin{equation}
\left\langle S_{12} \right\rangle_{\frac{1}{2}\otimes\frac{1}{2}\rightarrow S=1,l \neq 0}=
    \begin{cases}
        -\frac{2l}{2l+3}, &   J=l+1\\
       2,  & J=l\\
        -\frac{2(l+1)}{2l-1},  &  J=l-1.
    \end{cases}
\end{equation}

Value of $\left\langle S_{12} \right\rangle$ vanishes when $l=0$ and $S=0$. Moreover, these values are valid for the spin half particles.

\subsection{Decay width}
In addition to mass spectra, predictions of decay width are also important in hadron spectroscopy. Annihilation decays are very helpful in identifying conventional mesons and multiquark structures \cite{kwong}\cite{chaturvedi}. Quarkonium annihilation into light particles helps to learn about the strong fine structure constant and its role in understanding the quark forces.

 \subsubsection{Leptonic decays}

 $^{3}S_{1}$ and  $^{3}D_{1}$ states of a quarkonium can decay into a pair of leptons through a virtual photon. The leptonic decay rate can be estimated from the conventional Van-Royen-Weisskopf formula \cite{van}. This formula with first-order radiative corrections is given by \cite{kwong}\cite{brodsky},

 \begin{equation}
     \Gamma (n^{3}S_{1}\rightarrow e^{+}e^{-})=\frac{4q^{4}\alpha^{2}|R_{nS}(0)|^{2}}{M_{nS}^{2}}\left(1-\frac{16 \alpha_{s}}{3 \pi}\right).
 \end{equation}

Similarly,
\begin{equation}
     \Gamma (n^{3}D_{1}\rightarrow e^{+}e^{-})=\frac{25q^{2}\alpha^{2}|R^{''}_{nD}(0)|^{2}}{2m_{Q}^{4}M_{nD}^{2}}\left(1-\frac{16 \alpha_{s}}{3 \pi}\right).
 \end{equation}

Here, $q$ is the Coulomb charge of the quark ($e_{c}=\frac{2e}{3}$), $\alpha$ is the fine structure constant ($\alpha = \frac{1}{137}$), $\alpha_{s}$ ($\alpha_{s}(c \bar c)=0.318$) is strong coupling constant, $m_{Q}$ denotes the mass of quark ($m_{c}=1.32GeV$). $M_{nS}$ and $M_{nD}$ are masses of decaying quarkonia states. $R_{nS}(0)$ and $R_{nD}(0)$ are radial wave functions at zero for $S$ and $D$ states. $R^{''}_{nD}(0)$ is the second derivative of the radial wave function at zero.

\subsubsection{Two photon decay}
The decay of different charmonium states into photons is given by \cite{kwong},

  \begin{equation}
     \Gamma (n^{1}S_{0}\rightarrow \gamma\gamma)=\frac{3q^{4}\alpha^{2}|R_{nS}(0)|^{2}}{m_{Q}^{2}}\left(1-\frac{3.4\alpha_{s}}{ \pi}\right).
 \end{equation}
\begin{equation}
     \Gamma (n^{3}S_{1}\rightarrow \gamma\gamma \gamma)=\frac{4(\pi^{2}-9)q^{6}\alpha^{3}|R_{nS}(0)|^{2}}{3 \pi m_{Q}^{2}}\left(1-\frac{12.6\alpha_{s}}{ \pi}\right).
 \end{equation}
\begin{equation}
     \Gamma (n^{3}P_{0}\rightarrow \gamma\gamma)=\frac{27q^{4}\alpha^{2}|R^{'}_{nP}(0)|^{2}}{m_{Q}^{4}}\left(1+\frac{0.2\alpha_{s}}{ \pi}\right).
 \end{equation}
\begin{equation}
     \Gamma (n^{3}P_{2}\rightarrow \gamma\gamma)=\frac{36q^{4}\alpha^{2}|R^{'}_{nP}(0)|^{2}}{5m_{Q}^{4}}\left(1-\frac{16 \alpha_{s}}{3 \pi}\right).
 \end{equation}

Here, $R^{'}_{nP}(0)$ is the first derivative of the radial wave function at zero. All these formulae include QCD first-order radiative corrections.
% \vspace{-0.3cm}
%\vspace{-0.87cm}
 \subsubsection{Two gluon decay}
% \vspace{-0.47cm}

Experimental detection of di-gluon annihilation decay faces challenges because the gluonic state breaks down into multiple hadrons, which makes direct measurement difficult and first principle approximations unreliable. The decay width of a pseudoscalar meson via di-gluon annihilation, including the leading order QCD radiative correction, is given by \cite{kwong}\cite{lansberg},
 \begin{equation}
     \Gamma (n^{3}P_{0}\rightarrow gg)=\frac{6\alpha_{s}^{2}|R^{'}_{nP}(0)|^{2}}{m_{Q}^{4}}\left(1+\frac{9.5\alpha_{s}}{3 \pi}\right).
 \end{equation}

\begin{equation}
      \Gamma (n^{3}P_{2}\rightarrow gg)=\frac{8\alpha_{s}^{2}|R^{'}_{nP}(0)|^{2}}{5m_{Q}^{4}}\left(1-\frac{2.2\alpha_{s}}{3 \pi}\right).
 \end{equation}

\section{Results and Discussions} \label{discussion}

  The mass spectra of quarkonia ($c \bar c$) can be calculated using the formula,

\begin{equation}M_{c \bar c}=2m_{c}+E_{nl}+E_{spin}.
\end{equation}

The parameters used in the calculation to get charmonium masses are given in Table \ref{table:1}. 
%The results for the mass spectra for charmonium are given in Table \ref{table:2} to Table \ref{table:4}. The predicted mass spectra agreed well with experimental and other theoretical results. 
%The parameters are chosen to get a good fit for the experimental results. 

\begin{table}[H]
\addtolength{\tabcolsep}{-1pt}
%\begin{tablenote}
%\FloatBarrier
\caption[] {Parameters used in the calculation.}
%\end{tablenote}
\begin{center}
\label{table:1}
\begin{tabular}{llllll}
\hline
Parameters&$\alpha_{s}$ \cite{chaturvedi}&$b$ $(GeV)$ &$c$ $(GeV)$&$\mu$ $(GeV)$ &$m_{q}$ $(GeV)$\\
\hline
$c \bar c$&0.318&0.150& 1.00&0.03&1.32\\
\hline

%\FloatBarrier
\end{tabular}
\end{center}
\end{table}
%\end{landscape}

\subsection{Mass spectra of $c \bar c$}
Mass spectra for charmonium states are calculated and shown in Table \ref{table:2}, Table \ref{table:3} and Table \ref{table:4}. The results are compared with different theoretical results and available experimental results \cite{pdg}. Here, Chaturvedi and Rai \cite{chaturvedi}\cite{chaturvedi1} Sultan {\it et al,} \cite{sultan} and Soni {\it et al,} \cite{soni} used Cornell potential plus spin-dependent corrections. Ebert, Faustov, and Galkin \cite{ebert} used a relativistic quark model based on the quasipotential approach, and Kalinowski and Wagner \cite{kali} used the LQCD method. We have calculated the mass spectra of different states of charmonium, including $S$, $P$, and $D$ states. The spectroscopic notation for the states can be represented as $n^{2S+1}L_{J}$ (symbols have the usual meaning).

%\begin{landscape}
\begin{table}[H]
%\scalebox{0.9}{
%\begin{tablenote}
%\FloatBarrier
\caption[Mass spectra of charmonium ($S$- states) using Cornell potential] {Mass spectra of charmonium ($S$-  states) in $GeV$. }
%\end{tablenote}
\begin{center}
\label{table:2}
\scalebox{0.9}{
\begin{tabular}{lllllllll}
\hline
State &Present&PDG \cite{pdg}&\cite{chaturvedi}&\cite{chaturvedi1}& \cite{sultan}&\cite{soni}&\cite{ebert}& \cite{kali}\\
\hline
%&Cornell&Screen&logarithmicarithmic&Motyka&&&&&&&  \\
%\hline
%State    & P1      & P2     & P3    & P4     & PDG                                      & AR1   & AR2   & MS    & NR    & EB    & $L_QCD$ \\

$1^1S_0$ & 2.976   & 2.9839 $\pm$ 0.0004    & 3.004 & 2.989 & 2.982 & 2.989 & 2.981 & 2.884   \\
\hline
$1^3S_1$ &  3.082   & 3.096900 $\pm$ 0.000006 & 3.086 & 3.094 & 3.090 & 3.094 & 3.096 & 3.056   \\
\hline
$2^1S_0$ &  3.550  & 3.6375 $\pm$ 0.0011    & 3.645 & 3.572 & 3.630 & 3.602 & 3.630 & 3.535   \\
\hline
$2^3S_1$  & 3.642   & 3.68610 $\pm$  0.00006   & 3.708 & 3.649 & 3.672 & 3.681 & 3.672 & 3.662   \\
\hline
$3^1S_0$  & 4.027                                          & & 4.124 & 3.998 & 4.058 & -     & 4.043 &         \\
\hline
$3^3S_1$  & 4.111                                          & & 4.147 & 4.062 & 4.129 & -     & 4.072 &         \\
\hline
$4^1S_0$   & 4.445   &    & 4.534 & 4.372 & 4.384 & 4.448 & 4.384 &         \\
\hline
$4^3S_1$  & 4.523 &   & 4.579 & 4.428 & 4.406 & 4.514 & 4.406 &     \\    
\hline
%\FloatBarrier
\end{tabular}}
\end{center}
\end{table}
%\end{landscape}

\begin{table}[H]
%\begin{tablenote}
%\FloatBarrier
\caption[Mass spectra of charmonium ($P$- states) using Cornell potential] {Mass spectra of charmonium ($P$- states) in $GeV$. }
%\end{tablenote}
\begin{center}
\label{table:3}
\scalebox{0.9}{
\begin{tabular}{lllllllll}
\hline
State &Present&PDG \cite{pdg}&\cite{chaturvedi}&\cite{chaturvedi1}& \cite{sultan}&\cite{soni}&\cite{ebert}& \cite{kali}\\
\hline
 $1^3P_0$  & 3.435   & 3.4147 $\pm$ 0.0030   & 3.440 & 3.473 & 3.424 & 3.428 & 3.413 & 3.421   \\
\hline
$1^3P_1$   & 3.479  & 3.51067 $\pm$ 0.00005 & 3.492 & 3.506 & 3.505 & 3.468 & 3.511 & 3.480   \\
\hline
$1^3P_2$    & 3.498   & 3.55617 $\pm$ 0.00007 & 3.511 & 3.551 & 3.549 & 3.480 & 3.555 & 3.536   \\
\hline
$2^3P_0$    & 3.852   &                                    & 3.932 & 3.918 & 3.852 & 3.897 & 3.870 &         \\
\hline
$2^3P_1$   & 3.896   &    & 3.932 & 3.949 & 3.925 & 3.938 & 3.870 &         \\
\hline
$2^3P_2$   & 3.919   & 3.9225 $\pm$ 0.0010   & 4.007 & 4.002 & 3.965 & 3.943 & 3.949 & 4.066   \\
\hline
$3^3P_0$  & 4.283  &                                    & 4.394 & 4.306 & 4.202 & 4.296 & 4.301 &         \\
\hline
$3^3P_1$  & 4.326    &                                    & 4.401 & 4.336 & 4.271 & 4.338 & 4.319 &         \\
\hline
$3^3P_2$  & 4.350  &                                    & 4.427 & 4.392 & 4.309 & 4.358 & 4.350 &     \\    
\hline
\end{tabular}}
\end{center}
\end{table}
%\end{landscape}

\begin{table}[H]
%\begin{tablenote}
%\FloatBarrier
\caption[Mass spectra of charmonium ($D$- states) using Cornell potential] {Mass spectra of charmonium ($D$- states) in $GeV$. }
%\end{tablenote}
\begin{center}
\label{table:4}
\scalebox{1}{
\begin{tabular}{llllllll}
\hline
State &Present&PDG \cite{pdg}&\cite{chaturvedi}&\cite{chaturvedi1}& \cite{sultan}&\cite{soni}&\cite{ebert}\\
\hline

$1^3D_3$   & 3.706  &                                  & 3.789 & 3.806 & 3.805 & 3.775 & 3.813 \\
\hline
$1^3D_2$ & 3.711 & 3.8237 $\pm$ 0.0005 & 3.814 & 3.800 & 3.795 & 3.772 & 3.759 \\
\hline
$1^3D_1$   & 3.705   & 3.7737 $\pm$ 0.0004 & 3.815 & 3.785 & 3.783 & 3.775 & 3.783 \\
\hline
$2^3D_3$   & 4.153  &                                  & 4.273 & 4.206 & 4.165 & 4.176 & 4.220 \\
\hline
$2^3D_2$  & 4.160   &                                  & 4.248 & 4.203 & 4.158 & 4.188 & 4.190 \\
\hline
$2^3D_1$   & 4.158  & 4.1910 $\pm$ 0.0050 & 4.245 & 4.196 & 4.141 & 4.182 & 4.105 \\
\hline
$3^3D_3$   & 4.551  &                                  & 4.626 & 4.568 & 4.481 & 4.549 & 4.574 \\
\hline
$3^3D_2$  & 4.559  &                                  & 4.632 & 4.566 & 4.472 & 4.557 & 4.554 \\
\hline
$3^3D_1$ & 4.560   &                                  & 4.627 & 4.562 & 4.455 & 4.553 & 4.507 \\
\hline
%$3^{1}D_{2}$& & 4.629& 4.196&&\\
%\hline
%\FloatBarrier
\end{tabular}}
\end{center}
\end{table}

\begin{figure}[htbp]
\caption{Comparison of different potentials (without spin-dependent terms).}
\centerline{\includegraphics[width=9cm]{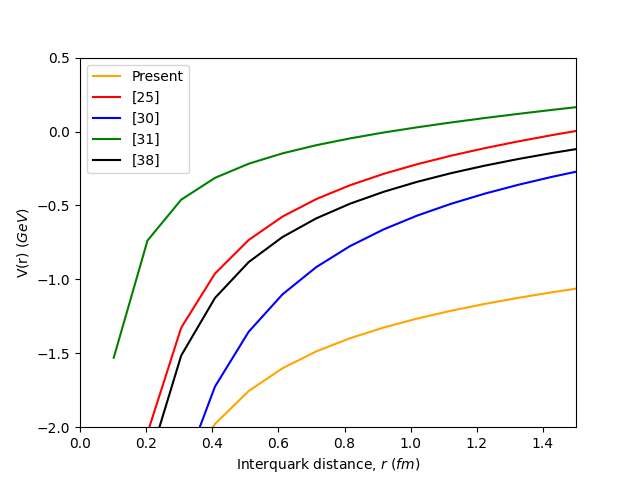}}
\label{fig: Figure 1}
\end{figure}

Table \ref{table:2}, Table \ref{table:3}, and Table \ref{table:4} show 
reasonable agreement with experimental data, while small deviations start 
to appear at increasing energy levels. Comparisons with other models, such 
as those from \cite{chaturvedi}, \cite{sultan}, and \cite{ebert}, indicate 
overall consistency, though differences are more pronounced for the  $3S$ 
and $4S$ states. Our results show better matching compared to other studies 
for $2^{3}P_{2}$ state in Table \ref{table:3}, whereas other studies 
overestimated the results. From Table \ref{table:4}, analysis of the $1^3D_1$, 
$1^3D_2$ and $1^3D_3$ states show correspondence between calculated results 
and theoretical expectations while showing small inconsistencies across 
different models. Experimental data from the $1^3D_2$ and $1^3D_1$ states 
show largest deviation with our findings compared to other models. In 
potential plots shown in Fig. \ref{fig: Figure 1}, the width of our proposed 
potential is more than the width of other potentials. Therefore, the energy 
eigenvalues of our proposed potential will be lowest among all these potentials. 
This may be the reason for the highest deviation among the compared ones.

\begin{table}[H]
%\begin{tablenote}
%\FloatBarrier
\caption[] {Mass splitting for $c \bar c$ ($MeV$).}
%\end{tablenote}
\begin{center}
\label{table:5}
\begin{tabular}{lllllllll}
\hline
State &Present&PDG \cite{pdg}&\cite{chaturvedi}&\cite{chaturvedi1}& \cite{sultan}&\cite{soni}&\cite{ebert}& \cite{kali}\\
\hline
$m_{J/\psi}-m_{\eta_{c}}(1S)$&106 &113 $\pm$0.7$\pm$0.1& 82&105&108&105&115&172\\
\hline
$m_{\psi}(2S)-m_{J/\psi}(1S)$&518&589.188$\pm$0.028& 704&660&690&692&691&778\\
\hline
%\FloatBarrier
\end{tabular}
\end{center}
\end{table}
%\end{landscape}

From Fig. \ref{fig: Figure 1}, it is clear that all these potentials have similar 
well-type structures. Therefore, their mass spectra will also show a similar nature. 
The energy gap among two consecutive energy levels decreases as we go for higher 
excited states because, for higher energy states, the width of the potential well 
gradually increases. Since all these potentials contain spin interaction terms, the 
energy gap among two consecutive energy levels does not maintain this behavior. 
This may be because of the significant role of the spin correction term in the mass 
spectrum of charmonium. From Tables \ref{table:2} to \ref{table:4}, this nature of 
the charmonium spectrum can be seen clearly. Especially $1D$ state data in Table 
\ref{table:4} for potential \cite{soni}. We have limited experimental data on
higher states. This highlights the need for further experimental efforts to 
establish the masses of higher states, which would help in the refinement and validation
of theoretical models of charmonium spectroscopy.

%When we 
%compare our results with the experimental results, our results show reasonably 
%good agreement which supports the screening effect included in our potential.

In the case of $S$ states, there is only spin-spin interaction, 
while $P$ and $D$ states have spin-orbit and tensor interactions. The mass splitting 
for different states of the same principal quantum number $n$ and orbital angular 
momentum $l$ is much smaller, especially for $P$ and $D$ states, indicating that 
these corrections are perturbative. The experimental study suggests $c \bar c$ 
\cite{pdg}, $m_{J/\psi}-m_{\eta_{c}}(1S)=113 \pm 0.7 \pm 0.1$ $MeV$. Similarly, 
$m_{\psi}(2S)-m_{J/\psi}(1S)=589.188 \pm 0.028$ $MeV$. The difference has been 
calculated and compared with experimental and other studies and given in Table 
\ref{table:5}. This way we find that overall our results show good agreement 
with the experimental results. Hence we can say that the screened potential is very 
close to the interquark potential.

\subsection{Decay width}

Different types of annihilation decay of charmonium states were calculated and compared with other studies. Experimental results are available for only a few decay modes. Calculated decay widths are shown in Table \ref{table:11} to Table \ref{table:13}. The results are compared with other studies. The constants, including quark masses, strong coupling constant, etc, are taken from the study of mass spectra, as mentioned earlier. 

\subsubsection{Leptonic decay ($nS_{1}^{3} \rightarrow e^{+}e^{-}$)}
\begin{table}[h!]
%\Floatbarrier
\caption[Leptonic decay width of $c \bar c$] {Leptonic decay width of $S$ states of $c \bar c$ ($keV$).}
\begin{center}
\label{table:11}
\begin{tabular}{llllllll}
\hline
State & Present&PDG \cite{pdg}&\cite{chaturvedi}&\cite{chaturvedi1}&\cite{soni}& \cite{vinod}\\
\hline
$1S_{1}^{3}$& 1.085&5.55$\pm$0.08&1.957& 6.932&2.925&5.47\\
\hline
$2S_{1}^{3}$& 0.967&2.33$\pm$0.01&1.178&3.727&1.533&2.14\\
\hline
$3S_{1}^{3}$& 0.529&0.86$\pm$0.01&0.969&2.994&1.091&0.796\\
\hline
$4S_{1}^{3}$&0.422&&0.860&2.638&0.856&0.288\\
\hline
\end{tabular}
\end{center}
\end{table}

For leptonic decay, experimental results are available for $1S$-$3S$ states. When the results for different leptonic decays are compared, it is observed that as the decays proceed to higher excited states, the decay widths become closer to experimental values. Decay rates given by Vinodkumar {\it et al} \cite{vinod} have shown good agreement with the experimental results. Vinodkumar used a relativistic model with harmonic oscillator potential. On the other hand, the decay rate given by Chaturvedi and Rai \cite{chaturvedi} and Soni {\it et al} \cite{soni} and our present calculation show 
good agreement with each other but a large difference from the experimental results.

Leptonic decay of charmonia can occur due to Quantum Electrodynamics (QED), weak 
interactions, or strong interactions. In charmonia, the quark annihilation may be due to all three interactions, but the lepton production will be due to QED and weak interactions only. Therefore, in charmonium decay, the QCD corrections become relevant to account for the interactions of quarks within hadrons. Chaturvedi and Rai \cite{chaturvedi} employed the NRQCD factorization expressions for various leading orders of decay. In another study, Chaturvedi and Rai \cite{chaturvedi1} applied radiative and quark propagator corrections, which appear to overestimate the values. The present study and study by Soni and others \cite{soni} used a similar type of QCD corrections, but the parameters were different and showed different results. Brodsky {\it et al} have shown that the running coupling constant can affect the perturbative expansions in QCD \cite{brodsky}. First-order corrections are comparatively small for most processes when the appropriate scale for the coupling constant is used. Our calculation shows a large difference from the experimental data. It indicates that in this formalism, we may have either missed some processes or some correction factors. Again, if we compare it with Vinodkumar's \cite{vinod} result, we find that relativistic model corrections are more successful than non-relativistic corrections for leptonic decays.

\subsubsection{Photonic decay ($nS_{0}^{1}$/$nP_{0}^{3}/nP_{2}^{3} \rightarrow \gamma \gamma$)}
\begin{table}[H]
\caption[Photonic decay width of $c \bar c$] {Photonic decay width for $c \bar c$ ($keV$).}
\begin{center}
\label{table:12}
\begin{tabular}{llllllllll}
\hline
State & Present&PDG \cite{pdg}&\cite{chaturvedi}&\cite{chaturvedi1}&\cite{soni}&\cite{lakhina}&\cite{gupta}&\cite{patel}\\
\hline
$1S_{0}^{1}$&6.289&& 6.725& 8.246&5.618 & 7.18& & \\
\hline
$2S_{0}^{1}$&5.839&&3.178& 4.560& 2.944& 1.71&&\\
\hline
$3S_{0}^{1}$&5.596&&1.493& 3.737&2.095 &1.21&&\\
\hline
$4S_{0}^{1}$&5.437&&0.858 &3.340&1.644&&&\\
\hline
$1P_{0}^{3}$& 8.752&2.341$\pm$0.189&4.185&2.692&&&6.38&7.33\\
\hline
$2P_{0}^{3}$&11.261&&4.306&4.716&&&&8.70\\
\hline
$3P_{0}^{3}$& 13.422&&4.847&8.078&&&&\\
\hline
$1P_{2}^{3}$& 1.095&&0.538&1.242&&&0.57&1.95\\
\hline
$2P_{2}^{3}$& 1.409&&0.554&1.485&&&&2.32\\
\hline
$3P_{2}^{3}$& 1.923&&0.626&1.691&&&&\\
\hline
%$2^{1}D_{2}$& & & &&\\
%\hline
%\FloatBarrier
\end{tabular}
\end{center}
\end{table}
%\end{comment}
For the di-photonic decay of charmonium, we have only one data point available. Our results show good agreement with Cornell potential \cite{chaturvedi}\cite{chaturvedi1}. However, with experimental data, a big variation is present. Here, we also find that there are huge differences among relativistic and non-relativistic data.

\begin{table}[H]
\caption[Tri-Photonic decay] {Tri-photonic decay width of $S$ states of $c \bar c$ ($eV$) ($nS_{0}^{1}$ $\rightarrow $ $\gamma \gamma \gamma$).}
\begin{center}
\label{table:a}
\begin{tabular}{llll}
\hline
State & Present& \cite{chaturvedi}&  \cite{chaturvedi1}\\
\hline
$1S_{0}^{1}$   & 1.046 &1.022 &2.997 \\
\hline
$2S_{0}^{1}$   & 0.972 & 0.900& 1.083\\
\hline
$3S_{0}^{1}$   & 0.931 & 0.857&1.046 \\
\hline
$4S_{0}^{1}$   & 0.905 &0.832 & 0.487\\
\hline
\end{tabular}
\end{center}
\end{table}
In the tri-photonic decay of charmonium, our results show good agreement with 
\cite{chaturvedi}. We do not have experimental data for this process, therefore, it 
will be too early to comment on it.

\subsubsection{Gluonic decay ($nP_{0}^{3}/nP_{2}^{3} \rightarrow gg$)}

\begin{table}[H]
\caption[Gluonic decay width of $c \bar c$] {Gluonic decay width of $c \bar c$ states ($MeV$).}
\begin{center}
\label{table:13}
\begin{tabular}{lllllll}
\hline
State & Present&\cite{chaturvedi1}&\cite{gupta}&\cite{patel}&\cite{bhaghyesh}\\
\hline
% &&&&&$(MeV)$&\\
%\hline
$1P_0^3$ & 25.18   & 14.19  & 13.44 & 32.58 & 3.337  \\ 
\hline
$2P_0^3$ & 32.40   & 24.973 &       & 38.70 & 2.060  \\ 
\hline
$3P_0^3$ & 38.61   & 33.876 &       &       &        \\ 
\hline
$1P_2^3$ & 4.708   & 2.914  & 1.2   & 3.38  & 0.784  \\ 
\hline
$2P_2^3$ & 6.059   & 5.099  &       & 4.01  & 0.504  \\ 
\hline
$3P_2^3$ & 7.221   & 6.867  &       &       &        \\
\hline
\end{tabular}
\end{center}
\end{table}
%\end{comment}
In gluonic decay, experimental detection is a big challenge. Unfortunately, we do not have experimental data for this decay mode. Our data shows good agreement with \cite{chaturvedi1}, where the authors have used the Cornell potential. Our results show a large difference with \cite{bhaghyesh} and \cite{gupta} results. They have used relativistic models. In this way, we find a significant difference between the results of relativistic and non-relativistic potentials as observed in the case of photonic decay.

\section{Conclusions} \label{conclusion}

 In the present work, we have found that the mass spectra of charmonium using a non-relativistic screened potential model with spin-dependent corrections show good agreement with the experiment and other non-relativistic models. For relativistic or semi-relativistic models, mass spectra show differences.  Comparative study on different potential models, lattice QCD, relativistic, and experimental results indicates that the proposed screened potential with spin-dependent elements is very close to the interquark potential (the exact form is still unknown).
 On the other hand, for decay spectra, relativistic models show good agreement with experimental data compared to the non-relativistic models. However, decay spectra show underestimated results with non-relativistic models, which indicates that some processes may have been missed. It needs a separate and detailed analysis which is beyond the scope of this work. 

\section*{Acknowledgements} SM is thankful to Manipal Academy of Higher Education (MAHE) for the financial support under Dr. T. M. A. Pai scholarship program.

\end{document}